\documentstyle[12pt,dina4p,epsfig]{article}
\title{
  \vskip-2cm
  {\baselineskip16pt
    \centerline{\normalsize \tt DESY 95-159 \hfill ISSN 0418-9833}
    \centerline{\normalsize \tt hep-ph/9508337 \hfill}
    \centerline{\normalsize \tt August 1995 \hfill}
  }
  \vskip2cm
  {\bf
    Dijet Cross Sections at O($\alpha\alpha_s^2$) in
    Photon-Proton Collisions
  }
  \author{
    {M.\ Klasen,  G.\ Kramer} \\
    {II. Institut f\"ur Theoretische Physik}\thanks
    {Supported by Bundesministerium f\"ur Forschung und
     Technologie, Bonn, Germany under Contract 05\,6HH93P(5) and
     EEC Program "Human Capital and Mobility" through Network
     "Physics at High Energy Colliders" under Contract
     CHRX-CT93-0357 (DG12 COMA)} \\
     {Universit\"at Hamburg} \\
     {D - 22761 Hamburg, Germany}
    }
  \date{}
}
\begin{document}
\maketitle
\vspace{3cm}
\begin{abstract}
\thispagestyle{empty}
We have calculated inclusive two-jet production in low $Q^2$
$ep$ collisions at O($\alpha\alpha_s^2$) superimposing direct and
resolved contributions. The results are compared with recent
experimental data from the ZEUS collaboration at HERA.
\end{abstract}
\newpage
\setcounter{page}{1}
\section{Introduction}
At HERA interactions between almost real photons and protons produce jets
at high transverse momentum \cite{xxx1}. Due to the large momentum scale
production cross sections should be calculable in perturbative QCD.
In leading order (LO), i.e.\ in O($\alpha\alpha_s$), jet production
proceeds through two distinct processes: (i) the virtual photon interacts
directly with a parton in the proton (Fig.\ 1a) (direct component) or
(ii) the virtual photon acts as a source of partons which collide
with the partons in the proton (Fig.\ 1b) (resolved component).
In the first case the full energy of the photon participates in the
interaction and the fraction of the photon momentum $x_\gamma$ involved
in the hard scattering process is equal to one. In the resolved process,
however, $x_\gamma$ is always less than one. Another distinction is that
in LO the final state in the resolved process includes a photon remnant
in addition to two jets and the proton remnant whereas in the direct
process the photon remnant is absent. \\

\begin{figure}[htbp]
 \begin{center}
  \begin{picture}(12,6)
   \epsfig{file=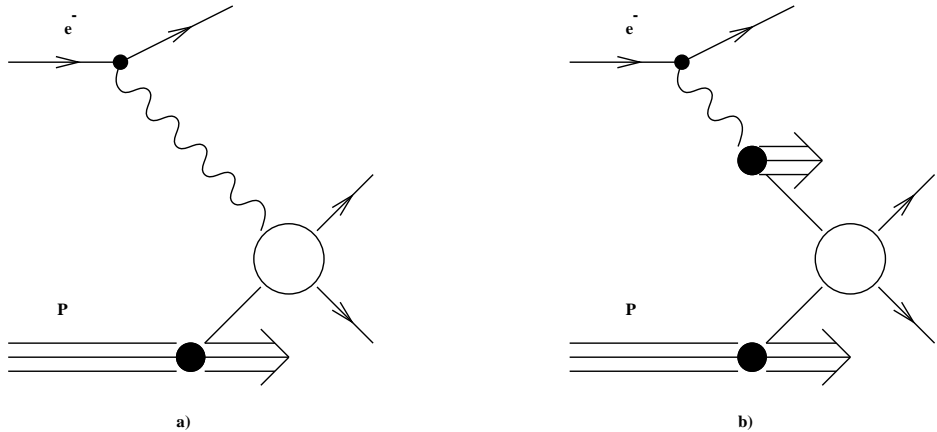,bbllx=160pt,bblly=90pt,bburx=450pt,bbury=700pt,%
           height=6cm,clip=,angle=-90}
  \end{picture}
  \caption{Schematic diagrams for a) direct and b) resolved photoproduction}
 \end{center}
\end{figure}

In next-to-leading order (NLO) of $\alpha_s$ this distinction between
direct and resolved contribution becomes ambiguous. Both components
are related to each other through the factorization scale at the photon
leg. The dependence of the NLO direct cross section on this scale
must cancel to a very large extent against the scale dependence in the
resolved cross section \cite{xxx2}. Higher order direct contributions
can have a photon remnant. Furthermore, $x_\gamma$ is no longer given
in terms of the kinematical variables of the dijet system. \\

Recently the ZEUS collaboration \cite{xxx3} published an analysis of
two-jet cross sections separating direct and resolved contributions with
a cut on $x_\gamma$. They measured the dijet cross section as a function
of the pseudorapidities of the two jets in regions which are sensitive to
the gluon momentum density in the proton for large $x_\gamma$ and
sensitive to the gluon density in the photon for small $x_\gamma$. This
procedure was suggested earlier by Forshaw and Roberts based on LO
QCD calculations \cite{xxx4} which have also been used in the analysis
of the ZEUS measurements. \\

Before information on the respective parton
densities can be gained from such an analysis several important effects
must be investigated in more detail. One of them is the influence of
NLO corrections to the direct and resolved photon cross sections on the
comparison of the data with the theory. Such calculations exist for
the inclusive single jet cross section for the direct \cite{xxx5,xxx6}
and resolved cross sections \cite{xxx5,xxx7} separately which have
been superimposed recently and compared to experimental data \cite{xxx8}.
B\"odeker also calculated the NLO direct inclusive two-jet cross section
as a function of the two-jet invariant mass and the two jet rapidities
\cite{xxx9}. For the comparison with the ZEUS data \cite{xxx3} the NLO
calculations must be performed in such a way that the experimental
constraints, in particular the cuts used to enrich the direct photon
contribution, can be built in easily. Such a calculation for direct
photoproduction in which the soft and collinear singularities of initial
and final state are isolated by an invariant mass resolution cut has
been completed recently by us \cite{xxx10}. Based on this work we have
calculated the NLO inclusive cross section for direct photoproduction
as a function of the average rapidity of the two jets and
of the transverse energy. We have built in the experimental constraints
on the kinematical variables as used in the ZEUS analysis. The contribution
of the resolved photon in the enriched direct $\gamma$ sample is
estimated in LO since it is supposed to contribute only a fraction in this
sample which is of interest for us. The effect of the NLO corrections
of the resolved photon to this sample is left for future work. \\

In section 2 we shall define our direct photon dijet cross section
and describe how it has been calculated. In section 3 we compare our
results with the ZEUS measurements and discuss their relevance towards
constraining the gluon distribution function of the proton at small $x$
and the structure function of the photon for large $x_\gamma$, also with
respect to the expectation of more accurate data in the future. We end
in section 4 with our conclusions. \\

\section{Direct Photon Dijet Cross Section}

We have calculated the inclusive cross section for two-jet photoproduction
coming from direct photons up to O($\alpha\alpha_s^2$) for final states
with at least two jets of $E_T > 6$ GeV. The photoproduction events
are defined by those $ep \rightarrow eX$ scattering events where the
$Q^2$ of the virtual photon is below $Q_{\max}^2 = 4~\mbox{GeV}^2$. We
describe the spectrum of the virtual photons by the Weizs\"acker-Williams
formula
\begin{equation}
 z F_{\gamma/e}(z) = \frac{\alpha}{2\pi}(1+(1-z)^2)\log\left(\frac{
                     Q_{\max}^2 (1-z)}{m_e^2~z^2}\right)
\end{equation}
where $z=E_\gamma / E_e$ is restricted to $0.2 < z < 0.8$ as
in the ZEUS analysis. $E_e = 26.7$ GeV and $E_p = 820$ GeV. \\

We adopt the jet definition of the snowmass meeting \cite{xxx11}
defining a jet as a collection of particles contained in a cone of radius
$R = 1$ in the plane of rapidity and azimuthal angle around the jet
momentum. This means that two partons may be considered as two separate
jets or as a single jet depending whether they lie outside or inside the
cone with radius $R$. In NLO the final state may consist of two jets
or three jets. Then the three-jet sample consists of all $2 \rightarrow 3$
parton scattering contributions which do not fulfill the cone condition. \\

The cross section calculated is d$\sigma$/d$\bar{\eta}$
where $\bar{\eta} = \frac{1}{2}(\eta_1+\eta_2)$ is the average rapidity of
the two jets with the requirement that the difference of the rapidities
$\eta^{\ast} = \eta_1 - \eta_2$ fulfills $|\eta^{\ast}| < 0.5$.
This cross section is the integral of
$\mbox{d}^3\sigma$/d$E_T$d$\bar{\eta}$d$\eta^{\ast}$ integrated over
$E_T > 6$ GeV and $-0.5 < \eta^{\ast} < 0.5$. The chosen $E_T$ is
the transverse energy of the so-called "trigger" jet with rapidity $\eta_1$
and $\eta_2$ is the rapidity of a second jet. The transverse energies
of these two jets fulfill $E_{T_1}, E_{T_2} > E_{T_3}$. We note that
$\bar{\eta}$ and $|\eta^{\ast}|$ are symmetric for $\eta_1 \leftrightarrow
\eta_2$. \\

For $2 \rightarrow 2$ parton scattering energy and momentum conservation
give the fraction of the photon energy participating in the hard
scattering as
\begin{equation}
  x_\gamma = \frac{E_T}{2 z E_e}\left( e^{-\eta_1}+ e^{-\eta_2}\right)
\end{equation}
where $z E_e$ is the initial photon energy and $\eta_1$ and $\eta_2$
are the rapidities of the two partons in the final state.
If there were exclusively two jets in the final state, $x_\gamma$ could
be determined from their kinematical variables. The events with $x_\gamma = 1$
are exclusively direct production and those with $x_\gamma < 1$
resolved production. In NLO more than two jets are produced and
$x_\gamma < 1$ is possible also for direct photoproduction. Therefore, in the
ZEUS dijet analysis the observable $x_\gamma^{\mbox{OBS}}$ was introduced,
which is defined as the fraction of the photon energy participating in
the production of the two highest $E_T$ jets with variables $E_{T_1}, \eta_1$
and $E_{T_2}, \eta_2$ respectively:
\begin{equation}
  x_\gamma^{\mbox{OBS}} = \frac{1}{2 z E_e}\left( E_{T_1} e^{-\eta_1}
  +E_{T_2} e^{-\eta_2}\right) .
\end{equation}
$x_\gamma^{\mbox{OBS}} \leq x_\gamma$ since the $x_\gamma$ in general
has contributions from all jets produced which means
\begin{equation}
  x_\gamma = \frac{1}{2 z E_e} \sum_{n} E_{T_n} e^{-\eta_n}
\end{equation}
where n runs up to $n = 3$ in NLO.
In the $x_\gamma^{\mbox{OBS}}$ distribution, nevertheless, we still expect
the direct and resolved processes to populate different regions since
the strictly two-jet samples of direct and resolved processes will
dominate the cross section. Therefore, the direct processes are concentrated
at large values of $x_\gamma^{\mbox{OBS}}$. The peak arising from the
direct contribution will not necessarily lie at $x_\gamma^{\mbox{OBS}} = 1$
due to higher order QCD effects. In the ZEUS analysis,
direct and resolved photoproduction events are separated by a cut
at $x_\gamma^{\mbox{OBS}} = 0.75$.
This value will also be used when we compare
our results with the ZEUS data. For the case of only two jets in the
final state we have $E_{T_1} = E_{T_2} \equiv E_T$, so that
$x_\gamma^{\mbox{OBS}}$ takes the following form when written in terms
of $\bar{\eta}$ and $\eta^{\ast}$:
\begin{equation}
  x_\gamma^{\mbox{OBS}} = \frac{E_T}{z E_e} e^{-\bar{\eta}} \cosh
  \frac{\eta^{\ast}}{2}.
\end{equation}
To select large $x_\gamma^{\mbox{OBS}}$ it is necessary to choose
$\bar{\eta} < 0$ for fixed $E_T$ and $\eta^{\ast} \simeq 0$. The
corresponding expression for the momentum fraction of the proton
entering the hard scattering process is
\begin{equation}
  x_p^{\mbox{OBS}} = \frac{E_T}{E_p} e^{\bar{\eta}} \cosh
  \frac{\eta^{\ast}}{2}
\end{equation}
so that $x_p^{\mbox{OBS}}$ can be small if $E_T$ is not too large.
So for $E_T = 6$ GeV and $x_\gamma = 1$ we have $x_p \simeq 2 \cdot 10^{-3}$.
Since the photon-gluon fusion $\gamma g \rightarrow q\bar{q}$ gives the
dominant contribution, one is able to probe the gluon structure function
of the proton to rather small values of $x$ \cite{xxx4}. \\

The theoretical framework of the calculation of the inclusive dijet
cross section is the same as in reference \cite{xxx10,xxx12}, where
further details can be found. To cancel infrared and collinear
singularities present in the $2 \rightarrow 3$ matrix elements and in
the virtual corrections to the $2 \rightarrow 2$ contributions, we apply
the phase space slicing method which has also been applied to the
calculation of jet production in $e^+ e^-$ collisions \cite{xxx13},
$\gamma p$ collisions \cite{xxx14} and in deep inelastic scattering
\cite{xxx15}. To separate the regions of phase space which contain
the singularities, we introduce an invariant mass cutoff $y$.
This cutoff is defined as usual with $s_{ij}/s < y$ where $s_{ij}$ denotes the
invariant mass squared of two particles $i$,$j$, and $s$ is the partonic
center of mass energy squared. Next we perform
partial fractioning to isolate infrared and collinear singularities.
Then we integrate the $2 \rightarrow 3$ cross sections over the phase
space region with soft and collinear singularities up to the invariant
mass cut. The remaining singularities which do not cancel among virtual
and real corrections are absorbed by the usual factorization and
renormalization into the parton densities of the photon and the proton.
For sufficiently small values of $y$ the relevant $2 \rightarrow 3$
subprocesses can be evaluated using an approximation where non-singular
terms are neglected in order to facilitate the analytical integration
over the soft and collinear regions of phase space. After this is
done the remainder of the 3-body phase space contains no singularities.
This procedure is particularly suited to build in kinematical constraints
as used in the analysis of the experimental data. \\

The further calculation now is based on two separate contributions -
a set of 2-body contributions and a set of 3-body contributions. Each set
consists of finite parts, all singularities have been cancelled or
subtracted and absorbed into structure functions. But each part
depends separately on the cutoff $y$. In case that experimentally 2-jet
and 3-jet cross sections could be measured with the same definitions
concerning the 2-jet and 3-jet part, they could be compared to these results
and the $y$ dependence of the different jet samples could be tested.
In this work we are interested in cross sections which are sufficiently
inclusive, and the separation with the invariant mass cutoff $y$ is
only a technical device. The dependence on $y$ must cancel in the inclusive
cross section. We checked this with the usual inclusive single-jet cross
section and a jet definition based on the cone algorithm described
above. Of course, the single-jet cross section now depends on the cone
radius $R$. We found complete independence of $y$ and perfect agreement
with the earlier results of B\"odeker \cite{xxx6} who used the
subtraction method to cancel soft and collinear singularities. To achieve
agreement we had to choose a rather small value of $y = 10^{-3}$.
Similar studies for jet photoproduction using different cutoffs were
performed by Baer et al. \cite{xxx14}. We also compared with the two-jet
invariant mass cross section in ref. \cite{xxx9} and found good agreement. \\

To be able to compare with the ZEUS measurements \cite{xxx3} we calculated
the inclusive two-jet cross section
d$\sigma$/d$E_T$d$\bar{\eta}$d$\eta^{\ast}$ for
$|\eta^{\ast}| < 0.5$ and integrated over $E_T > 6$ GeV where $E_T$ is
the transverse energy of the "trigger" jet. The result as a function
of $\bar{\eta}$ is shown in Fig. 2. In these curves no cut on
$x_\gamma^{\mbox{OBS}}$ is applied. The cross sections are for all
direct contributions in LO (where $x_\gamma^{\mbox{OBS}} = 1$) and in NLO
with no additional constraints except those on $\eta^{\ast}$ and $E_T$.
As structure function for the proton we have chosen CTEQ3M \cite{xxx16}, which
is a NLO parametrization with $\overline{\mbox{MS}}$ factorization and
$\Lambda^{(4)} = 238$ MeV. This $\Lambda$ value is also used to calculate
the two-loop $\alpha_s$ value at the scale $\mu = E_T$. The factorization
scale is chosen $M = E_T$ also.
In Fig. 2 we show three curves. The full curve is the LO cross section with
a maximum near $\bar{\eta} = 0$ and which is around 1.2 nb. The sharp
drop-off near $\bar{\eta} = 0$ is caused by the constraints on $z$ and on
$E_T$. The other two curves are NLO results with $y = 10^{-3}$. The
dashed curve is the genuine two-jet cross section containing the LO
contribution and all terms of the
three-parton final states with two partons having
momenta in the cone with radius $R=1$.
This cross section is negative due to $y$ dependent terms
($\log y$ is dominant) which
originate from the separation of the initial state singularities. Therefore,
$y$ acts as a physical cutoff which separates 2-jet contributions,
where one of the partons is recombined with the remnant jets, from the
genuine 3-jet contribution. Of course, when $y$ takes a more physical
(larger) value for this remnant recombination, the two-jet cross section
becomes positive. The dotted curve shows the contribution of the
3-jet final state, which is positive and much larger than the LO
prediction. This cross section depends also strongly on $y$ due to the
initial state singularities. The sum of 2-jet and 3-jet cross section
gives 1.5 nb in the maximum and is independent of $y$, i.e. it is
sufficiently inclusive to guarantee that the $y$ dependence drops out. \\

\begin{figure}[htbp]
 \begin{center}
  \begin{picture}(12,8)
   \epsfig{file=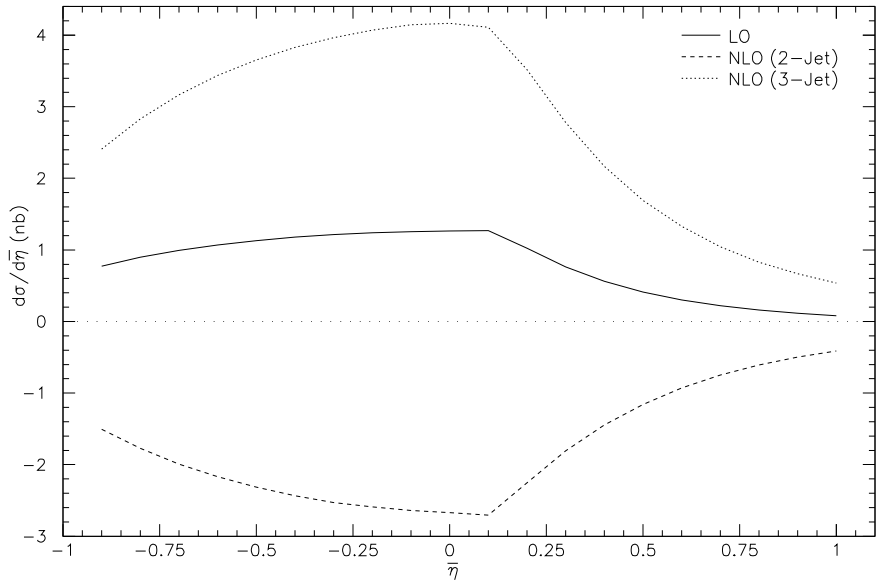,bbllx=100pt,bblly=90pt,bburx=520pt,bbury=710pt,%
           height=8cm,clip=,angle=-90}
  \end{picture}
  \caption{d$\sigma$/d$\bar{\eta}$ for $ep \rightarrow eX + 2$ (or more)
           jets for direct photoproduction with $E_T > 6$ GeV, $R = 1$
           as a function of $\bar{\eta}$. The full curve is the LO cross
           section, the dashed (dotted) curve is the NLO 2-jet (3-jet)
           cross section with invariant mass cut $y = 10^{-3}$.}
 \end{center}
\end{figure}

\section{Comparison with ZEUS Data}

Before we compare with the dijet cross sections as measured in the ZEUS
experiment \cite{xxx3} we investigate the influence of the additional
cuts on the NLO prediction. In these calculations we have taken
$y = 10^{-3}$ in order to be independent on the approximations used
in the analytical calculations.
In the analysis of the ZEUS measurements the additional requirements
are $x_\gamma^{\mbox{OBS}} \geq 0.75$ to enhance the direct contributions
and the constraint $E_{T_1}, E_{T_2} > 6$ GeV, i.e. the considered events
contain at least two jets with equal minimal transverse energy. These
two constraints influence the 3-jet cross section (dotted curve in
Fig. 2) but not the 2-jet cross section (dashed curve in Fig. 2). Then
immediately the problem arises whether the inclusive cross sections with
these cuts are independent of $y$, i.e. are "infrared safe". Unfortunately,
this is not the case due to $y$ dependence of the cross sections
originating from the initial state collinear singularities. \\

The cut
$x_\gamma^{\mbox{OBS}} \geq 0.75$ modifies only the 3-jet contribution,
which contains the 3-parton terms outside the cone radius $R$.
Furthermore, the corrections to the 2-jet cross section originating
from the initial state singularity at the photon leg have the same
structure as a resolved cross section and have contributions for all
$x_\gamma$, where $x_\gamma$ is the fraction of the photon energy
involved in the hard parton-parton scattering. If we separate the
$x_\gamma \geq 0.75$ terms in this contribution, i.e. subtract the
$x_\gamma < 0.75$ terms from the 2-jet cross section (dashed curve in
Fig. 2), we obtain a NLO correction which is independent of $y$ and
coincides with the prediction with no cuts on $x_\gamma$. The result
is shown as the dashed curve in Fig. 3. \\

\begin{figure}[htbp]
 \begin{center}
  \begin{picture}(12,8)
   \epsfig{file=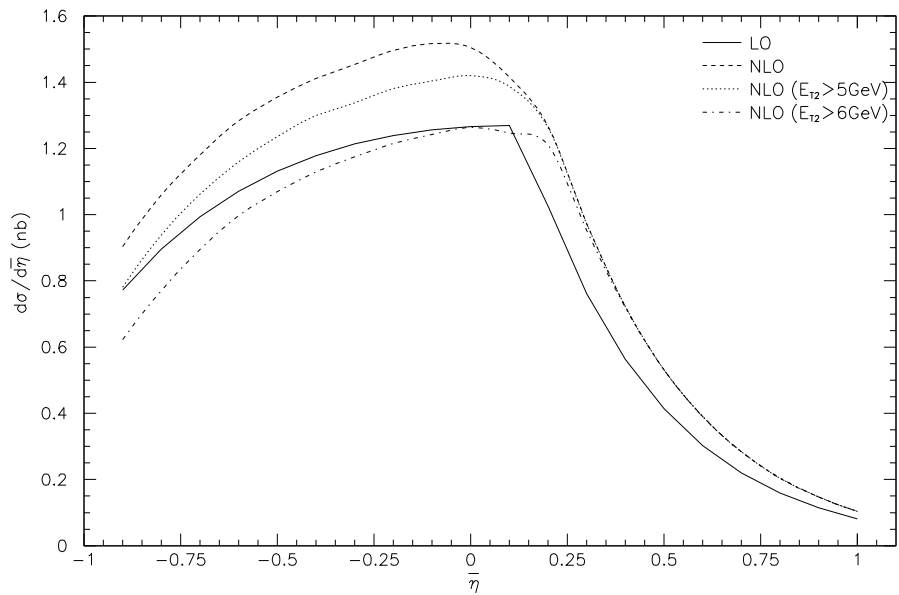,bbllx=100pt,bblly=90pt,bburx=520pt,bbury=710pt,%
           height=8cm,clip=,angle=-90}
  \end{picture}
  \caption{d$\sigma$/d$\bar{\eta}$ for NLO direct photoproduction with
           different constraints for the 3-jet contribution as a function of
           $\bar{\eta}$. The dashed curve is NLO from Fig. 2 with no
           constraints. The dotted (dot-dashed) curve is with $E_{T_1} > 6$
           GeV, $E_{T_2} > 5$ GeV ($E_{T_1}, E_{T_2} > 6~\mbox{GeV},~E_{T_3}
           \stackrel{<}{\scriptstyle>} 1$ GeV concerning 2- and 3-jet
           separation).}
 \end{center}
\end{figure}

The experimental cut $E_{T_1},
E_{T_2} > 6$ GeV is more problematic since it also leads to cutoff dependent
cross sections. With this strict cut on the
$E_T$ of both jets, there remains in some events very little transverse
energy for the third jet, so that the $y$ cut acts as a physical cut.
If we arrange the 2- and 3-jet contributions in such a way that we introduce
a cut on $E_{T_3}$, the transverse energy of the third jet, we obtain
$y$ cut independent NLO contributions again. We have introduced
$E_{T_3} = 1$ GeV as a physical cut, where the contribution with
$E_{T_3} < 1$ GeV is included in the two-jet cross section and the contribution
$E_{T_3} > 1$ GeV is included in the three-jet cross section.
With this
additional constraint on the 3-jet part of the inclusive cross section
we can demand $E_{T_1}, E_{T_2} > 6$ GeV. The result is the dot-dashed
curve in Fig. 3, which we consider our final answer for the NLO
direct cross section incorporating the ZEUS cuts.
We see that it is almost the same as our LO prediction. This means that due
to the various cuts the NLO correction is reduced very much.
If we require only $E_{T_1} > 6$ GeV, $E_{T_2} > 5$ GeV we also obtain
a $y$-cut independent result. Then, the third jet can have enough transverse
energy to compensate the initial state singularities.
The cross section is 10\% larger (dotted curve in Fig. 3) in this case. \\

The results in Fig. 3 do not include any tails,
i.e. $x_\gamma \geq 0.75$ contributions, from
the resolved cross section. This contribution has been estimated in
\cite{xxx3} with the LO photon structure function of Gordon and Storrow
\cite{xxx17} taking $E_T/2$ as the factorization scale and $\Lambda = 200$ MeV.
This leads to a contribution of 0.15 nb in the maximum of
d$\sigma$/d$\bar{\eta}$ for the resolved tail. To be consistent
with our NLO calculations, we must choose a $\overline{\mbox{MS}}$ NLO
photon structure function. Then the resolved contribution for
$x_\gamma \geq 0.75$ is much larger. In Fig. 4 we plotted this
resolved cross section as a function of $\bar{\eta}$ for the photon
structure functions GRV($\overline{\mbox{MS}}$),
GRV($\mbox{DIS}_\gamma$) \cite{xxx18},
GS(HO), which is NLO in the $\overline{\mbox{MS}}$ scheme, and
GS(LO, Set 2) \cite{xxx17},
which was considered in \cite{xxx3}, but now with scale $M_\gamma = E_T$
instead of $M_\gamma = E_T/2$ as used in the ZEUS analysis.
Already this scale change increases the resolved contribution with
GS(LO, Set 2)
by 70\%. In other words, the resolved cross section in the region
$x_\gamma \geq 0.75$ is highly uncertain and can change by a factor of two,
if different NLO photon structure functions are considered. \\

\begin{figure}[htbp]
 \begin{center}
  \begin{picture}(12,8)
   \epsfig{file=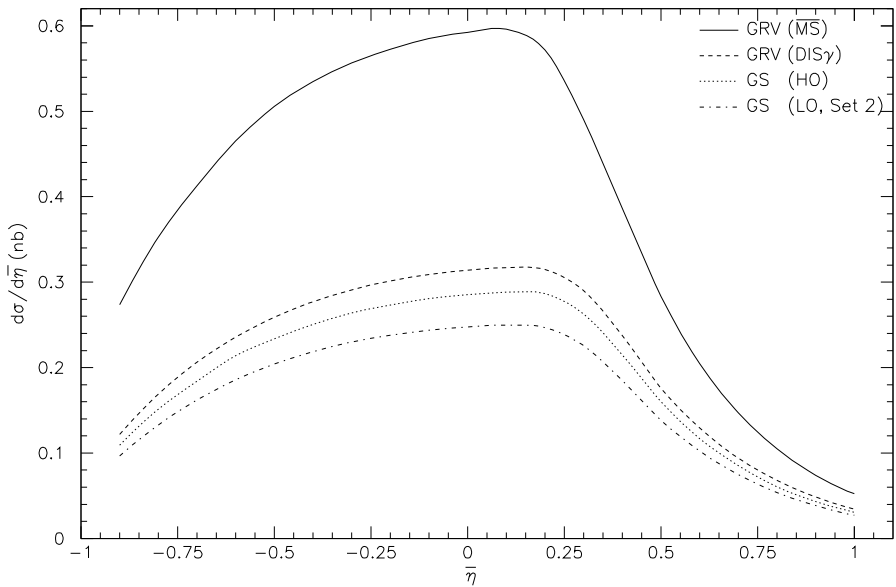,bbllx=100pt,bblly=90pt,bburx=520pt,bbury=710pt,%
           height=8cm,clip=,angle=-90}
  \end{picture}
  \caption{d$\sigma$/d$\bar{\eta}$ for LO resolved photoproduction using
           different photon structure function parametrizations as a
           function of $\bar{\eta}$.}
 \end{center}
\end{figure}

In Fig. 5 we added the resolved cross sections with the
GRV($\overline{\mbox{MS}}$) and the GS(HO) structure functions to the NLO
direct contribution from Fig. 3. The experimental points are from the ZEUS
analysis where we corrected their cross sections for hadronization
effects taken from their Fig. 3c. In addition, we added the energy
scale uncertainty in quadrature to the combined statistical and
systematic error. In Fig. 5 we see that our prediction combined with the
NLO GS(HO) photon structure function reproduces the data very well.
The prediction with the GRV($\overline{\mbox{MS}}$) structure function
is mostly above the data by a factor approximately 1.2, and it coincides
approximately with the LO curve, which has been obtained by adding also
the resolved contribution with the GRV($\overline{\mbox{MS}}$) structure
function. This reflects the fact that in Fig. 3 the NLO direct cross
section with all the cuts was not very different from the LO curve
except for additional contributions in the $\bar{\eta} > 0$ region
and a reduction in the $\bar{\eta} < 0$ region.
We repeated the NLO calculation with the GRV photon structure function
in the $\mbox{DIS}_\gamma$ scheme, which is by a factor of two smaller than
the GRV function in the $\overline{\mbox{MS}}$ scheme (see Fig. 4).
For consistency, the subtracted terms in the photon structure function
appear in the NLO direct cross section. Therefore the sum is not changed
when we convert to the $\mbox{DIS}_\gamma$ scheme \cite{xxx2}.
This has been checked explicitly, i.e. the NLO GRV curve is for the
$\overline{\mbox{MS}}$ and the $\mbox{DIS}_\gamma$ scheme. For this
consistency it was essential that the resolved cross section which contains
no NLO terms in the hard scattering is calculated with NLO proton and
photon structure functions and with the same $\alpha_s$, i.e. in two loops
and with the same $\Lambda$ value. LO structure functions and $\alpha_s$
in one-loop would also spoil the $M_\gamma$ scale compensation between
the NLO direct and resolved contributions. \\

\begin{figure}[htbp]
 \begin{center}
  \begin{picture}(12,8)
   \epsfig{file=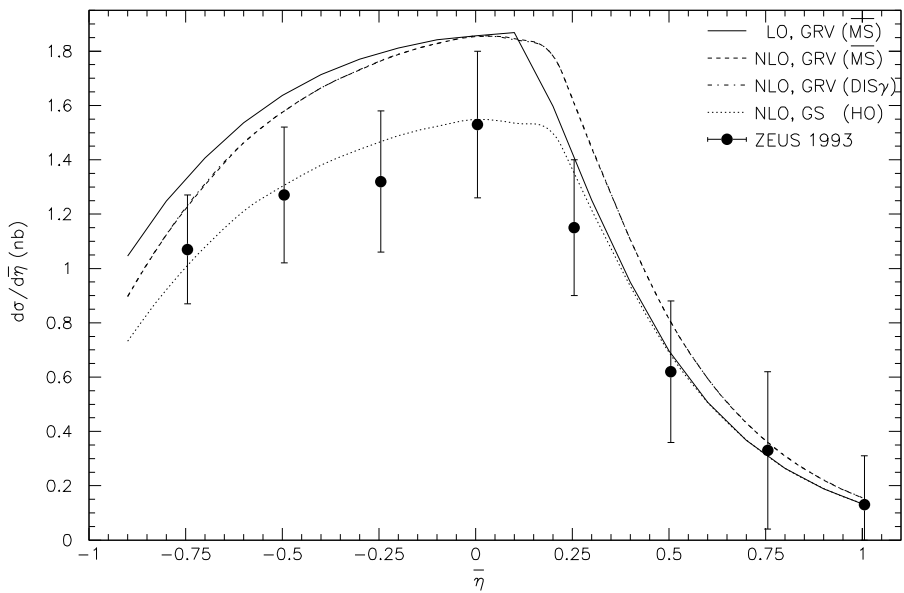,bbllx=100pt,bblly=90pt,bburx=520pt,bbury=710pt,%
           height=8cm,clip=,angle=-90}
  \end{picture}
  \caption{Sum of NLO direct and LO resolved cross sections
           d$\sigma$/d$\bar{\eta}$ as a function of $\bar{\eta}$
           compared to data of ref. [3]. Four curves for different
           photon structure functions are shown: LO (full curve), NLO
           with GRV ($\overline{\mbox{MS}}$ and $\mbox{DIS}_\gamma$)
           (dashed and dot-dashed), and NLO with GS (HO) (dotted).}
 \end{center}
\end{figure} From our analysis it is obvious that the resolved contribution to
the two-jet cross section for $x_\gamma^{\mbox{OBS}} \geq 0.75$ is not small.
It can amount up to 50\% of the direct contribution depending on the
photon structure function. It appears that in the region
$x_\gamma \geq 0.75$ the quark distribution of the
GRV($\overline{\mbox{MS}}$) structure function is much larger than for the
GS(HO). The gluon part is small in this region. Indeed, for $Q^2 = 10~
\mbox{GeV}^2$ the difference is approximately 50\% \cite{xxx23}. At larger
$Q^2$, which is relevant for our calculations, the difference must be
even larger (see Fig. 5, compare with Fig. 4). Therefore, the inclusive
two-jet cross section in the large $x_\gamma$ region is suitable to obtain
information on the quark distribution in the photon. By changing the
boundary of the $x_\gamma$ region towards $x_\gamma \rightarrow 1$ one
might be able to establish whether the NLO quark distributions in the
photon have the rather singular behaviour towards $x_\gamma \rightarrow 1$
as predicted by the GRV or AFG functions \cite{xxx19},
or if they behave more like
in the GS structure function. We remark that a similar difference
occurs in the LO GRV and GS structure functions \cite{xxx23}. But both
change appreciably for $x_\gamma^{\mbox{OBS}} \geq 0.75$,
if one applies the NLO versions. \\

The two-jet cross section changes very little ($\simeq 0.1$ nb) when
we use other proton structure functions which produce the existing deep
inelastic data as well as CTEQ3M, as for example MRS(A'), MRS(G)
\cite{xxx20},or
GRV($\overline{\mbox{MS}}$) \cite{xxx21}. This means, these structure
functions all have more or less the same gluon distribution, which
is very much determined by deep inelastic data and other data used in the
analysis. Therefore, the direct part of the two-jet cross section is
very well predicted by our NLO calculation and the emphasis is more on
the resolved part in the large $x_\gamma$ region. \\

Here our results are only an estimate since the NLO corrections are not
included in the resolved cross section. From calculations of the
inclusive one-jet cross sections it is known that these corrections for
$R = 1$ are large taking LO results with NLO structure functions and
two-loop $\alpha_s$ as the basis. This would enlarge the discrepancy
of the predictions in Fig. 5 with the existing data even more for the
GRV choice and may also lead to disagreement
for the GS(HO) photon structure function. \\

Besides NLO corrections for the resolved part, there are other important
topics which need further studies: the influence of possible jet pedestal
energies on the data, the influence of hadronization and of incoming
parton transverse momentum. The latter has been investigated for ingoing
gluons in the direct cross section \cite{xxx22}. Its effect seems
to decrease the direct cross section by approximately 0.3 nb \cite{xxx3},
which might compensate for the larger resolved cross section with
the GRV structure function in Fig. 4 and 5. \\

\section{Conclusions}

Differential dijet cross sections d$\sigma$/d$\bar{\eta}$ have been
calculated in NLO for the direct and in LO for the resolved part as a
function of $\bar{\eta}$. The kinematical constraints $|\eta^{\ast}| < 0.5$,
$E_T > 6$ GeV, $0.2 < z < 0.8$, $Q_{\max}^2 = 4~\mbox{GeV}^2$,
$x_\gamma^{\mbox{OBS}} \geq 0.75$, and $R = 1$ have
been incorporated as in the ZEUS experiment. The cross sections have
been obtained with the phase space slicing method for cancelling soft and
collinear divergences in NLO. The infrared stability and the independence
of the factorization scheme and scale at the photon leg have been tested.
It turns out that the final result depends very much on the choice of the
photon structure function for the resolved part. We think that the
direct part is reliably predicted in NLO. Therefore measurements of the
dijet cross sections in the large $x_\gamma$ region offer the
possibility to get information on the quark distribution of the photon
near $x_\gamma = 1$. \\
\newpage

\end{document}